\documentclass[a4paper,11pt]{article}

\usepackage{amsmath}
\usepackage{amstext}
\usepackage{graphicx}
\usepackage{hyperref}
\newcommand{\meinehomepage}{\href{http://btm8x5.mat.uni-bayreuth.de/~bothmer/}
                     {http://btm8x5.mat.uni-bayreuth.de/$\tilde{\,\,}$bothmer}
                           }
\newcommand{\includepicture}[3]
                       {\begin{center}
                        \includegraphics[#2]{#1}
                        \end{center}
                       }
\begin{document}
\title{Predicting critical crashes? A new restriction for the free variables.}
\author{Hans-Christian Graf v. 
        Bothmer\footnote{Laboratoire J. A. Dieudonne,
                         Universit\'e de Nice Sophia-Antipolis,
                         Parc Valrose,
                         F-06108 Nice Cedex 2,
                         bothmer@web.de} \,
        and
        Christian Meister\footnote{H\"olderlin Anlage 1
                                   D-95447 Bayreuth,
                                   christian.meister@eurocopter.com}
        }

\maketitle

\parindent 0cm
\parskip 0.2cm

\section{Introduction}

Several authors have noticed the signature of log-periodic oscillations
prior to large stock market crashes \cite{sornette1996}, 
\cite{feigenbaumfreud1996}, \cite{vandewalle1998}. 
Unfortunately good fits of the corresponding equation to stock market prices 
are also observed in quiet times. To refine the method several approaches 
have been suggested:

\begin{itemize}
\item Logarithmic Divergence: Regard the limit where
 the critical exponent $\beta$ converges to 0. \cite{vandewalle1998}
\item Universality: Define typical ranges for the free parameters,
by observing the best fit for historic crashes. \cite{criticalcrashes}
\end{itemize}

We suggest a new approach. From the observation that the hazard-rate
in \cite{criticalcrashes} has to be a positive number, we get an inequality
among the free variables of the equation for stock-market prices.

Checking 88 years of Dow-Jones-Data for best fits, we find that $25\%$ of
those that satisfy our inequality, occur less than one year before
a crash. We compare this with other methods of crash prediction, i.p.
the universality method of Johansen et al., which followed by a crash only in
$9\%$ of the cases.

Combining the two approaches we obtain a method whose predictions
are followed by crashes in $54\%$ of the cases.  

\section{The hazard rate}
\label{hazardrate}

In \cite{criticalcrashes} Johansen et al suggest, that during a
speculative bubble the crash hazard rate
$h(t)$, i.e, the probability per unit time that the crash will happen
in the next instant if it has not happened yet, can be modeled by
by
\[
 h(t) \approx B_0(t_c-t)^{-\alpha} 
              + B_1(t_c-t)^{-\alpha}\cos(\omega\log(t_c-t)+\psi).
\]
By assuming that the evolution of the price during a speculative
bubble satisfies the martingale (no free lunch) condition, they obtain
a differential equation for the price $p(t)$ whose solution is
\[
   \log \left( \frac{p(t)}{p(t_0)} \right)
   = \kappa \int_{t_0}^t h(t')dt'
\]
before the crash. Here $\kappa$ denotes the expected size of the crash.

This implies that the evolution of the logarithm of the 
price before the crash
and before the critical date $t_c$ is given by:
\[
  (*) \quad  \log(p(t)) \approx 
                         p_c 
                         - \frac{\kappa}{\beta}B_0(t_c-t)^\beta 
                         - \frac{\kappa}{\sqrt{\beta^2+\omega^2}} 
                           B_1(t_c-t)^\beta \cos(\omega \log(t_c-t) + \phi)
\]
With $\beta=1-\alpha$, $p_c$ the price at the critical date,
and $\phi$ a different phase constant.

Now the hazard rate is a probability and therefore positive. This leads
to a necessary condition:
\begin{align*}
  &    &  0 &\le h(t)\\
  &\iff&  0 & \le  B_0(t_c-t)^{-\alpha} 
                   + B_1(t_c-t)^{-\alpha}\cos(\omega\log(t_c-t)+\psi) \\
  &\iff&  0 & \le  B_0+ B_1\cos(\omega\log(t_c-t)+\psi) 
\end{align*}
since $t < t_c$. At some times near the critical date 
$\cos(\omega\log(t_c-t)+\psi)$ takes on the values $-1$ and $1$. 
This implies the necessary conditions
\[
   0  \le  B_0\pm B_1 \iff |B_1| \le B_0. 
\]
On the other hand these conditions are also sufficient for $h(t) \ge 0$ since
$\cos(\omega\log(t_c-t)+\psi)$ is always between $-1$ and $1$.

To summarize, if the assumptions of Johansen et al are valid, we must
have $|B_1| \le B_0$ prior to a critical crash.

\section{88 Years of Dow Jones}

To check this model of speculative bubbles, we have investigated 
the Dow Jones index from $1912$ to $2000$. This period contains $23668$
trading days. 

Johansen et al define
a crash as a continuous drawdown (several consecutive days of negative index 
performance) larger than $15\%$. The following diagram shows the drawdowns 
of the Dow Jones index from $1912$ to $2000$. Observe that there have been 
$4$ drawdowns larger than $15\%$ namely in the years $1929$, $1932$, 
$1933$ and $1987$.

%\begin{center}
%\epsfig{file=down-kurz.png, width=13cm}
%\end{center}

\includepicture{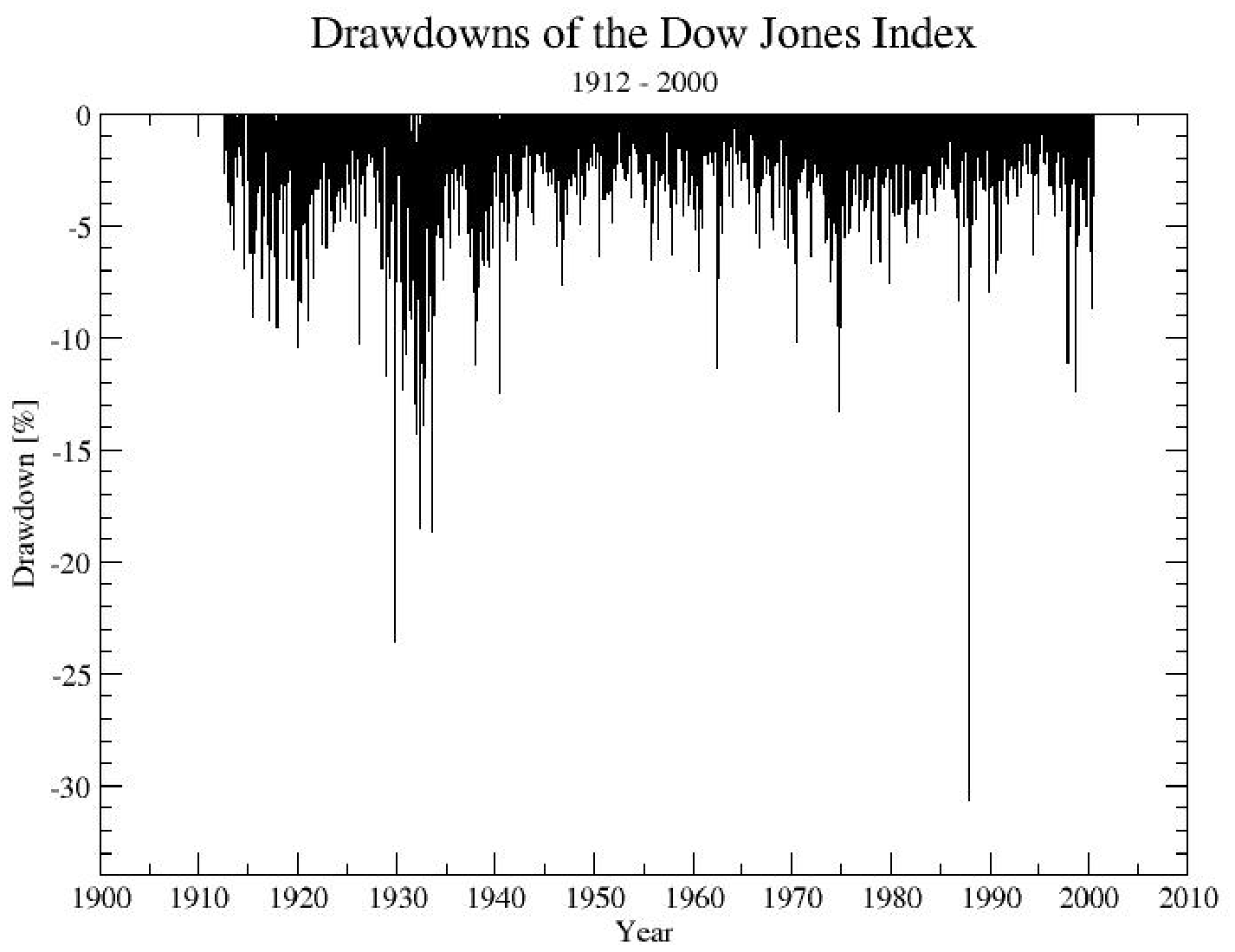}{width=14cm}{.pdf}

As our basic data set, we have calculated numerically 
the best fit of equation $(*)$ to a sliding window of $750$ trading 
days, every $5$ trading days. This yields $4761$ best fits. The complete
data set is available at \meinehomepage.

In what follows, we will call a crash prediction 
successful, if it was issued at most
one year before a crash. With this definition there are $229$ best fits
that could possibly give a successful crash prediction. By predicting
crashes randomly, one would obtain a successful prediction in $4.8\%$
of the cases.

\subsection{Mean square errors $\chi$}

The first approach to detecting speculative bubbles is to look for
good fits of $(*)$ to the Dow Jones Index. If the mean square error
$\chi$ of the fit is sufficiently small one issues a crash prediction. 

The next figure shows the mean square errors of all our best fits compared
with the best fits before a crash. Unfortunately small errors also occur
in quiet times. 

\includepicture{chi_vorcrash}{width=13cm}{}

If one issues a crash prediction if the mean square error
is smaller then $0.75$ one obtains:

\begin{center}
\begin{tabular}{|c|c|c|}
\hline
                  & before crash & not before crash \\
\hline
no crash prediction ($\chi \ge 0.75$)   & $175$        & $2799$ \\
\hline
crash prediction ($\chi < 0.75$)        & $72$         & $1732$\\
\hline
\end{tabular}
\end{center}

I.e. only $72/(72+1732)\approx 3.9\%$ of the predictions are successfull.
Since this is worse than issuing random predictions, we conclude
that one can not predict a crash by looking only at the mean square 
error. This observation has also been made by Sornette and Johansen.

\subsection{Critical Times $t_c$}

If the model of Johansen et al is correct one should expect that
a crash occurs close to the critical date $t_c$. Using this
one can issue a crash prediction when the critical date $t_c$ is 
less than one year away. Using our dataset this lead to:

\begin{center}
\begin{tabular}{|c|c|c|}
\hline
                  & before crash & not before crash \\
\hline
no crash prediction ($t_c\ge today + 1\, year$)    & $93$        & $2652$ \\
\hline
crash prediction ($t_c < today + 1\, year$)        & $136$       & $1879$\\
\hline
\end{tabular}
\end{center}

I.e $136/(136+1879) \approx 6.7\%$ of the predictions are successful. This is
slightly better that random predictions, but still not very good.

\subsection{Universality}
\label{universality}

Johansen et al suggest that speculative bubbles exhibit universal
behavior. This would imply that $\beta$ and $\omega$
take on roughly the same values for each speculative bubble. The following
diagram shows the distribution of $\omega$ before crashes compared
with the distribution during other times. 

\includepicture{omega}{width=13cm}{}

One can clearly observe
an unexpected peak around $\omega = 9$ before the crashes. If we
issue a crash prediction in the range $7 < \omega < 13$ we obtain:

\begin{center}
\begin{tabular}{|c|c|c|}
\hline
                  & before crash & not before crash \\
\hline
no crash prediction                    & $150$     & $3728$ \\
\hline
crash prediction ($7 < \omega < 13$)   & $79$      & $803$\\
\hline
\end{tabular}
\end{center}
 
I.e $79/(79+803) \approx 8.9\%$ of the predictions are successful. 

The distribution of $\beta$ before crashes and not before crashes is:

\includepicture{beta}{width=13cm}{}

Here we observe a tendency toward lower values
of $\beta$, but no clear peak. We interpret this as evidence, that one
should look for logarithmic divergence, i.e. the limit of $\beta$ 
tending to $0$, as suggested by Vandewalle et al. \cite{vandewalle1998}.
We will investigate this approach in a later paper.

\subsection{Positive hazard rate}

In section \ref{hazardrate} we have explained that in the 
model of Johansen et al. 
the hazard rate $h(t)$ must be positive. We proved that this is
equivalent to 
\[
     |B_1| \le B_0.
\]
From our best fits we can calculate the value
\[
       \kappa(B_0 - |B_1|),
\]
which should also be positive during a speculative bubble, since
$\kappa$ is a positive number. Consequently we can issue a crash warning
if $\kappa(B_0 - |B_1|)$ is positive. With our dataset we obtain:

\begin{center}
\begin{tabular}{|c|c|c|}
\hline
                  & before crash & not before crash \\
\hline
no crash prediction ($\kappa(B_0 - |B_1|)\le0$)  & $133$     & $4255$ \\
\hline
crash prediction ($\kappa(B_0 - |B_1|)>0$)   & $96$      & $276$\\
\hline
\end{tabular}
\end{center}

I.e $96/(96+276) \approx 25,8\%$ of the predictions are successful.
This is already a practical success rate, but we can do even better, if
we combine this with universality.

\subsection{Positive hazard rate and universality}

Combining the last two approaches we issue a crash prediction,
if the hazard rate is everywhere positive and $\omega$ is in
the range of section \ref{universality}. This gives

\begin{center}
\begin{tabular}{|c|c|c|}
\hline
                  & before crash & not before crash \\
\hline
no crash prediction   & $164$     & $4476$ \\
\hline
crash prediction 
                      & $65$      & $55$\\
($\kappa(B_0 - |B_1|)>0$ and $7 < \omega < 13$)    & &\\
\hline
\end{tabular}
\end{center}

I.e $65/(65+55) \approx 54.1\%$ of the crash predictions are successfull.

The following diagram shows when these crash predictions where issued.
For every trading day we have plotted the number
 of crash predictions during the
past year and the drawdown of the Dow Jones index. 

\includepicture{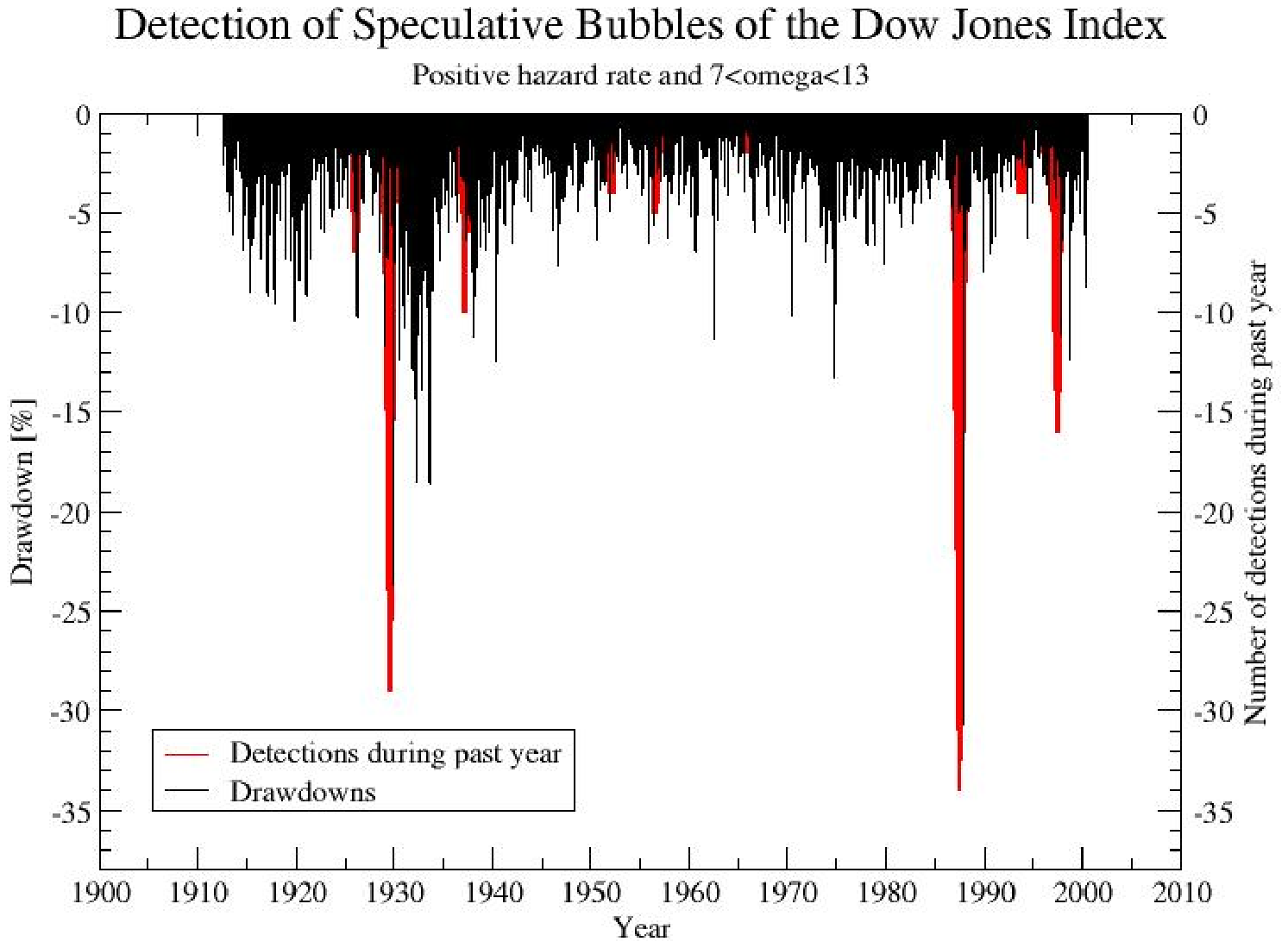}{width=13cm}{.pdf}

%\begin{center}
%\epsfig{file=down-kurz+indicator+omega.png, width=13cm}
%\end{center}

Notice that the crashes
of $1929$ and $1987$ have been predicted well in advance. The crashes
of $1932$ and $1933$ have not been directly predicted, but we argue that
they are in the aftermath of $1929$ and represent the bursting of
the same speculative bubble. The crash predictions of $1997$ where followed
by two small crashes in October $1997$ and $1998$, which didn't quite
reach $15\%$. One could argue that they represent a crash in two steps.

\section{Summary}

We have derived a new restriction of the free variables in
the model of Johansen et al \cite{criticalcrashes} for 
stock market prices during a speculative 
bubble. This restriction alone yields crash predictions with 
a $25\%$ successrate for the Dow Jones index. 
This is an improvement over the $9\%$ successrate
obtained by using universality. Combining our approach and
the universality method we obtain a success rate of $54\%$.

We think that these results represent strong evidence for the model
of Johansen et al describing speculative bubbles in the stock market.

\end{document}